\documentclass[twocolumn]{openjournal}
\usepackage{color}
\usepackage{graphicx, amsmath, amsmath, amssymb}
\usepackage[breaklinks, colorlinks, citecolor=blue]{hyperref}
\usepackage{listings} 

\usepackage[T1]{fontenc}

\newenvironment{mytabular}[1][1]{
  \renewcommand*{\arraystretch}{#1}
  \tabular
}{
  \endtabular
}

\shorttitle{CESs and crossing angles}
\shortauthors{Thomas, McCallum, Brown}

\begin{document}
\title{Consequences of constant elevation scans for instrumental systematics in Cosmic Microwave Background Experiments}
\author{Daniel B. Thomas}
\email{dan.b.thomas1@gmail.com}
\affil{School of Physics and Astronomy, Queen Mary University of London, London, E1 4NS, UK}
\author{Nialh McCallum}
\author{Michael L. Brown}
\affil{Jodrell Bank Centre for Astrophysics, School of Physics \& Astronomy, The University of Manchester, Manchester M13 9PL, UK }
\date{\today}

 \begin{abstract}
 Instrumental systematics need to be controlled to high precision for upcoming Cosmic Microwave Background (CMB)  experiments. The level of contamination caused by these systematics is often linked to the scan strategy, and scan strategies for satellite experiments can significantly mitigate these systematics. However, no detailed study has been performed for ground-based experiments. Here we show that under the assumption of constant elevation scans (CESs),  the ability of the scan strategy to mitigate these systematics is strongly limited, irrespective of the detailed structure of the scan strategy. We calculate typical values and maps of the quantities coupling the scan to the systematics, and show how these quantities vary with the choice of observing elevations. These values and maps can be used to calculate and forecast the magnitude of different instrumental systematics without requiring detailed scan strategy simulations. As a reference point, we show that inclusion of even a single boresight rotation angle significantly improves over sky rotation alone for mitigating these systematics. A standard metric for evaluating cross-linking is related to one of the parameters studied in this work, so a corollary of our work is that the cross-linking will suffer from the same CES limitations and therefore upcoming CMB surveys will unavoidably have poorly cross-linked regions if they use CESs, regardless of detailed scheduling choices. Our results are also relevant for non-CMB surveys that perform constant elevation scans and may have scan-coupled systematics, such as intensity mapping surveys.
 \end{abstract}

\maketitle

\section{Introduction}
Ground-based Cosmic Microwave Background (CMB) surveys are a crucial part of ongoing (e.g. \cite{bicep2015,polarbear2019,sptpol2019,actmetric}) and future (e.g. \cite{somain,cmbs4}) attempts to map the CMB temperature and polarisation in exquisite detail.

Systematics, including instrumental systematics, need to be controlled to high precision in order to achieve the science goals of these surveys (see e.g. \cite{hu2003}), and various techniques are being investigated for mitigating and controlling different systematics (see e.g. \cite{bicepsyst,Wallis_map,QE}). Some systematics, particularly instrumental systematics such as differential gain and pointing errors between differenced detectors, couple to the scan strategy (see e.g. \cite{hu2003,odea2007,shimon1,shimon2,bock2009,wallis,hivon2017, formalism}). This coupling means that knowledge of the scan strategy is required to compute the expected magnitude of the contamination from these systematics, and also that a careful choice of scan strategy may be able to mitigate these systematics. For so-called ``ideal\footnote{Scans where each sky pixel is visited at a large number of regularly spaced angles.}'' scans, the contamination from these systematics is reduced to zero. \cite{wallis} showed that satellite scans can be constructed to reduce the contamination from these systematics by 2 to 3 orders of magnitude. However, no such general study has been performed for ground based experiments. In particular, it is not clear how to optimise a ground-based scan strategy to get close to an ideal scan, and how close to an ideal scan can be reached in principle.

The scan strategies of ground based surveys are complicated and hard to parameterise. As a result, detailed simulations of these scan strategies require many involved choices and considerations, such as: how and when to scan different parts of the sky; how long to scan each patch for before moving to a different patch of sky; when and how to avoid the Sun and Moon; what range of azimuth ``throw'' to use and at what speed to scan in azimuth; which elevations to scan at; how to equalise depth over different parts of the sky; and which times of the year to scan. The simulations themselves can be computationally expensive, depending on the level of realism and detail in the simulations. Combined with the high number of parameters and choices, this expense can make searches over scan strategy parameter space prohibitive. In this paper, we will refer to the full set of choices, parameters, times and durations that are required to completely specify a ground-based scan strategy, as the ``detailed scheduling choices''.

The majority of ground based surveys are composed of constant elevation scans (CESs), which involves choosing an observing elevation, and holding the elevation fixed at this value whilst sweeping the telescope in azimuth. CESs are used because they are good for instrument stability, and because varying the elevation during a scan affects the column length of atmosphere through which the telescope is looking, resulting in more complicated noise and systematic properties. As a result, the elevation is typically held fixed whilst the instrument is scanning, even if multiple different elevations are used for scanning at different times. The use of CESs puts strong constraints on the possible properties of the survey, particularly in regards to how the scans couple to contamination from instrumental systematics.

In this work we show that the assumption of CESs allows one to put bounds on the parameters that couple the scans to the instrumental systematics, and calculate typical values and maps for these parameters, \textit{independently of detailed scheduling choices}. In particular, we show that there is a fundamental limit to how close any scan strategy composed of CESs can get to an ideal scan for the purposes of these parameters. These bounds and typical values allow forecasts, estimations and calculations to be carried out without making detailed scheduling choices, and these bounds can also be used as a target when optimising the detailed scheduling choices according to the science goals of a particular experiment. These results also allow the relevant features of detailed scans to be approximated with simple simulations.

The parameters describing the coupling of instrumental systematics to the scan strategy are closely related to the standard metric for the cross-linking \citep{actmetric}. Cross-linking refers to the different paths that the instrument takes across the sky as it scans, which can be important for mitigating the effects of 1/f atmospheric noise during the map-making step (see e.g. \cite{0807.3658,polarbearmap}). Whilst we focus on scan-coupled systematics, we also comment on the relevance of our findings for any surveys whose science goals require good cross-linking, which plan to use CESs. We reiterate that these findings are independent of detailed scheduling choices.

We describe our approach and code implementation in section \ref{sec_setup} and we present the consequences of CESs in section \ref{sec_scans}. In section \ref{sec_skyrot} we use the limitations of CESs to examine the crossing-angle coverage achievable for surveys based in the Atacama Desert, and the resulting bounds on the degree to which scan-coupled systematics can be mitigated. We present our conclusions in section \ref{sec_conc}.

\section{Scan-coupled systematics and ground-based scanning}
\label{sec_setup}
Many instrumental systematics have a fixed orientation in the focal plane of the telescope, particularly for polarisation sensitive detectors.
These systematics include pointing errors, anisotropic beam effects, and gain or calibration effects for polarisation detectors. Since the orientation of the focal plane with respect to the sky changes according to how the instrument scans the sky, the contamination caused by these systematics depends on the scan strategy \citep{wallis,formalism}.

The dependence of the contamination on the scan strategy can be computed within each pixel using the parameters $h_k$, defined as 
\begin{equation}
    h_{k} =\frac{1}{N_{\text{hits}}} \sum_{j} e^{ik\psi_j}=\frac{1}{N_{\text{hits}}} \sum_{j} \left( \cos(k\psi_j)+i\sin(k\psi_j)\right)\text{,}
    \label{eqn_hn}
\end{equation}
where $k$ takes integer values. The scan information being used here is the set of crossing angles $\psi_j$, which give the relative orientations of the focal plane to the sky for each measurement within the pixel in question. We show how to calculate these angles below. Note that similar parameters for examining crossing angle coverage are used in \cite{hu2003,odea2007,shimon1,shimon2,bock2009,hivon2017}.

The $|h_k|^2$ parameters measure the variation within a set of crossing angles\footnote{Note that one can add a constant value to every angle in a set without changing the value of $|h_k|^2$ for that set.}. The value of $|h_k|^2$ (for any $k$) ranges from $0$ to $1$, where $0$ is the value for an ideal scan and $1$ represents the worst possible scan (every crossing angle in the pixel is the same). For satellite scan strategies, values as small as $|h_k|^2\sim10^{-3}$ can be reached by careful optimisation \citep{wallis}. There has previously been no detailed investigation for ground based surveys, in particular one that examines: which decisions and choices in the complicated ground-based scan strategies are important for determining the $h_k$ values in each pixel; what typical values of $h_k$ are for ground-based surveys; how well ground-based surveys can be optimised in this regard; and how close ground-based surveys can get to an ideal scan for the purposes of these parameters.

These $h_k$ values are important for several reasons. Different $h_k$ are required to calculate the contamination due to different systematics. For example, temperature to polarisation leakage due to differential pointing errors within differenced detector pairs couples to $h_1$ and $h_3$ \citep{wallis,formalism}, and leakage caused by differential ellipticity effects couples to $h_2$ and $h_4$. We refer the reader to section 2 of \cite{formalism} for general expressions showing how signals of different spin couple to different $h_k$, and to equations (26)-(28) of \cite{wallis} for some explicit examples of how the $h_k$ parameters manifest in the power spectrum contamination for specific systematics. We note that these parameters are also relevant when detector-differencing is not used \citep{formalism}. In addition, insufficiently accurate incorporation of the scan strategy (i.e. $|h_k|^2$ maps) into the contamination caused by systematics can result in different conclusions for quadratic estimators that are used to find and clean systematics from the maps \citep{QE}. Computing $h_k$ maps is also critical for constructing simple and fast map-based simulations that incorporate the effects of systematics, notably systematics that can usually only be studied with full time ordered data (TOD) simulations \citep{formalism}.

In this work we focus on $k=1$ to $4$ as the most relevant for intensity to polarisation leakage (see e.g. \cite{wallis}), although we note that there do exist systematics that couple to $h_k$ with $k>4$ \citep{formalism}. The same approach taken here can be used to study these additional parameters if required. 

Since the $h_k$ parameters capture general information about the crossing angles, they relate to other aspects of scan strategies beyond the coupling of particular systematics. For example, the $h_2$ and $h_4$ quantities appear directly in simple binning polarisation map-making. More interestingly, in the recent ACT release \citep{actmetric} a metric was used to quantify cross-linking, which corresponds to $\sqrt{|h_2|^2}$ (in the absence of boresight rotation; see below and section \ref{sec_instrotate}). Due to this correspondence, a corollary to our results is that the constraints on $|h_2|^2$ constitute a limitation on the achievable levels of cross-linking (if using CESs) that is independent of detailed scheduling choices. Such a limitation is likely to impact the use of maximum likelihood map makers. This is a complicated issue beyond the scope of this work so we do not explore it here, other than noting that the maps and values of $|h_2|^2$ later in this paper facilitate such a study without requiring detailed scheduling choices.

The connection between $|h_2|^2$ and cross-linking breaks down in more complicated scenarios where different types of crossing angles need to be tracked. For example, when there is boresight rotation \citep{bicepsyst} or rotating half wave plates that modulate the incoming photon polarisation (see e.g. \cite{Maino02,Brown09,0611394,0710.0375,1510.01771,1709.04842}), neither of which improves cross-linking. Here we primarily focus on the simple case with only sky rotation, and briefly comment on boresight rotation in section \ref{sec_instrotate}. Half wave plates require tracking the crossing angles and half wave plate angles separately, resulting in a more complicated set of $h_{k,l}$ parameters (see \cite{wallis_hwp} for details), so we do not consider them further in this work.

There are ways to remove these systematics such as during map-making. This requires knowledge of the spins of the systematics and which $h_k$ they couple to (see e.g. \cite{Wallis_map,formalism,im}. In this work we focus on understanding the range of possible values and map structures of the $h_k$ for ground-based scan strategies, and which aspects of the scan strategies these parameters depend on.

\subsection{Calculating crossing angles in ground-based scans}
To calculate the possible $h_k$ values and maps for ground based surveys, we need to calculate the possible crossing angles $\psi_j$ that can be achieved within each pixel. To do this calculation, we make extensive use of the python pyEphem \citep{pyephem} and Healpy\footnote{http://healpix.sourceforge.net} \citep{healpy1,healpy2} packages.

We define the observatory using an \texttt{ephem.Observer()} object, where, for these results, the key property of the observatory is its Latitude. We take the right ascension (RA), and declination (Dec), of the centre of a healpix pixel as an \texttt{ephem.FixedBody()} object. The FixedBody \texttt{Compute} method is used to compute the elevation and azimuth\footnote{For the purposes of this work we will define azimuth as increasing clockwise from North, such that a ``rising'' field has an azimuth ($\theta$) of $0^\circ<\theta<180^\circ$.} required to see the healpix pixel from the given observatory at a particular time. This chain of computational steps is the basic building block of the code.
 
We then want to calculate the orientation of the instrument focal plane on the sky. Since we are restricting to CESs, we know that the next observation will occur at the same elevation, at a slightly increased azimuth value $\delta \theta $.\footnote{Note that we are interested in the angle of the focal plane with respect to the sky, which means that we cannot apply equation \ref{eqn_calcpsi} for both increasing and decreasing azimuth scans without applying an additional correction to one of the two cases, as the opposite edges of the focal plane are the leading edges in the two situations. Correctly accounting for this effect results in the same crossing angle at the instant that a measurement is taken, regardless of whether azimuth is increasing or decreasing. For simplicity in our code, and without loss of generality, we always treat $\delta \theta$ as positive.} The RA and Dec corresponding to the new azimuth can then be calculated using the \texttt{radec\_of} method in pyEphem. Given the RA and Dec of the pixel, and the RA and Dec of the next observation, the orientation of the focal plane with respect to the sky can be calculated using the two-argument arctangent as
\begin{eqnarray}
&&\delta \text{Dec}=\text{Dec}_2-\text{Dec}_1\\
&&\delta \text{RA}=\text{RA}_2-\text{RA}_1\\
&&\psi=\arctan2\left(\delta \text{Dec},\delta \text{RA} \cos(\text{Dec}_1)\right)\label{eqn_calcpsi}\text{.}
\end{eqnarray}
Appropriate choices of $\delta \theta$ depend on the scan speed and sampling frequency of the telescope. For this work we use $\delta \theta=0.8'$, corresponding to for example a scan speed of $0.4$ degrees per second and a sampling frequency of $30$Hz. However, our results are robust to an order of magnitude change either side of our chosen value.\footnote{This means that the crossing angle results derived here also apply for scan strategies with varying azimuth scan speed, including during the turnaround periods where the telescope switches from an ``East to West'' scan to a ``West to East'' scan, as long as the (varying) azimuth scan speed is within this range.}

Combining the above, we have a routine that computes the crossing angle $\psi$ (and the required elevation and azimuth for this observation) as a function of the observatory Latitude, the time, and the RA and Dec of the targeted pixel. We use this routine in several ways to generate the results in the rest of the paper. In particular, we use this routine to sample the possible crossing angles of a given pixel over a fixed period of time (from a single day to 5 years) given a range of possible observing elevations. We also use the routine to fix the elevation, and then generate all possible crossing angles within each pixel that can be achieved using that elevation.

By setting up the problem as above, we have made the implicit assumption that every hit in a scan will be at the centre of the corresponding map pixel. This assumption introduces a small error, but for the pixel sizes typical of current surveys, the declination variation within a pixel is small enough for this to be negligible (we will see in the next section that the declination variation within each pixel is the only relevant aspect of this assumption).

In this work we are interested in understanding the limits of what can be achieved in principle, so the setup we use assumes that each element of the focal plane is free to scan the sky independently. In practice, different focal plane elements do not have this freedom and effectively scan at slightly different elevations to the nominal (boresight) elevation. This constraint is one of the practical reasons that the bounds we derive are unlikely to be reachable in practice, even when the bounding value is relatively poor.

We note that balloons performing constant elevation scans during circumpolar flights at an approximately constant latitude will also be well modelled by our approach, however if there is significant latitude variation amongst the parts of the balloon flight during which data is taken, then the approach taken here would need to be generalised in order to describe this situation.

\section{Limitations of CES\lowercase{s} and consequences for scan-coupled systematics}
\label{sec_scans}
In this section we use the codes described earlier to examine what the assumption of CESs means for ground-based surveys. The first subsection is not new, but it is a simple and self-contained summary of the relevant aspects of crossing angle coverage, which is all easily obtained from our codes. This summary is required to understand how we construct the limits on $h_k$ variation that we present in later subsections.

\subsection{CES implications for crossing angles}
Using the simple simulation tools developed above, we explore the parameter space that could affect the crossing angle of a given observation. The parameters we varied were RA; Dec; observing elevation; time, year and season of observation; rising vs setting (azimuth less than or greater than $180^\circ$); and telescope latitude and longitude. We summarise the key findings here. 
\begin{enumerate}
\item During a 24-hour period, for a given observing elevation, the point on the sky in question can be observed at two times, each with a different azimuth; one rising and one setting.
\item Each of these gives a single distinct crossing angle, so for $N$ distinct observing elevations, there is a maximum of $2N$ distinct crossing angles per sky pixel.
\item The rising and setting crossing angles have the same magnitude but opposite signs.
\item For fixed declination, the largest and smallest (rising) crossing angles correspond to the largest and smallest observing elevations. The variation with elevation is monotonic.
\item For fixed elevation, there is a monotonic dependence of (rising) crossing angle on declination.
\item The variation of crossing angle with elevation is such that, for fixed latitude, fields with Dec$\sim$latitude (fields which pass close to overhead), have the smallest range of possible crossing angles.
\item The variation in crossing angle from day to day, over the course of a year, or from year to year, is negligible.
\item The RA and telescope longitude are both unimportant.
\end{enumerate}

\begin{figure}
  \centering
 \includegraphics[width=\columnwidth]{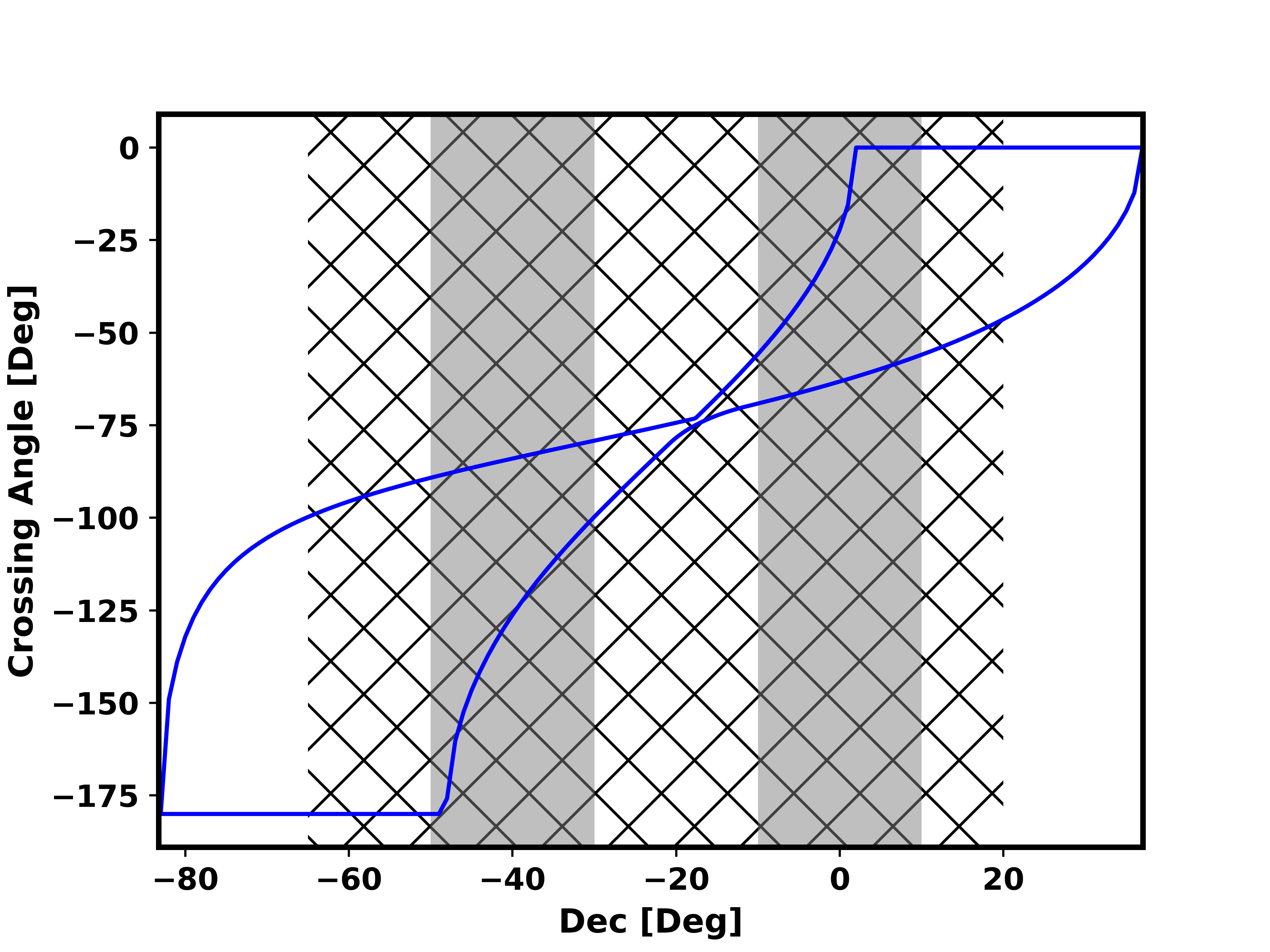}
\caption{The crossing angles as a function of Declination, for SO location and elevation bounds $30^\circ$ and $65^\circ$, are shown by the two blue lines. The range of crossing angles possible for each declination corresponds to the space between the two lines at each point. The large hatched region corresponds to the approximate range of declinations that will be targeted by the SO Wide survey, and the two darker shaded areas correspond to the approximate ranges of declinations that will be targeted by the SO Deep surveys.}
\label{fig_psidec}
\end{figure}

We illustrate these results in figure \ref{fig_psidec}, where we show the range of crossing angles available as a function of declination, for a telescope located at Latitude $-22^\circ 56.396'$. This latitude corresponds to CMB experiments in the Atacama Desert, such as the Simons Observatory (SO) site. The blue lines correspond to the crossing angles obtained from nominal minimum and maximum observing elevations of $30^\circ$ and $65^\circ$ respectively. The range of crossing angles possible for each declination corresponds to the space between the two lines at each point. The narrowing of the range of crossing angles for fields with Dec$\sim$latitude is clearly visible. The hatched region corresponds to the approximate range of declination that will be targeted by the SO Wide survey \citep{soscan}, and the two darker shaded areas correspond to the approximate ranges of declination that will be targeted by the SO North and South Deep surveys \citep{soscan}. We note that the deep scans avoid the range of declinations where the crossing angle coverage is worst. At the same time, it is clear from the figure that their placement in declination is not optimal from the point of view of maximising the crossing angle coverage. The Wide survey has an unavoidable region of poor crossing angle coverage, that \textit{cannot} be mitigated by detailed scheduling choices without removing the assumption of CESs.

\subsection{Optimising $|h_k|^2$ in each pixel with CESs}
\label{sec_2ers}
From the previous section, we see that for a fixed observatory location and targeted declination, the only relevant variables for calculating the possible values of $h_k$ within a pixel at a given sky location are the bounding two elevations at which observations are taken, and whether we wish to include observations that are rising, setting or both. Given these minimal pieces of information, we can calculate $h_k$ in that pixel for \textit{every} possible scan strategy composed of CESs by sampling the appropriate number of values from the range of crossing angles bounded by the observing elevations, with appropriate sign choices according to whether they are rising or setting. The set of possible pixel values calculated according to this procedure shows the range of possible pixel values that could be achieved by any ground-based experiment. In particular, as long as the assumptions of the previous section hold, most notably that pixels are small enough, no detailed scheduling choices can deliver a result outside of this range within that pixel.

We know that the worst possible value of $h_k$ in any pixel is $1$, corresponding to only observing that pixel with a single crossing angle. This is possible within the context of CESs, so to establish the range of possible $h_k$ values within each pixel, we only need to find the optimal (lowest) value of $h_k$ within each pixel. This amounts to finding the best set of $\psi_j$ between the two endpoints (corresponding to the observing elevation bounds), such that the magnitudes of $\frac{1}{N}\sum^N_j \cos \left(k\psi_j\right)$ and $\frac{1}{N}\sum^N_j \sin \left(k\psi_j\right)$ are minimised.

For $k=1$ to $4$, the elevation range $30^\circ-65^\circ$, and the SO latitude, the $h_k$ metrics are optimised (minimised) for nearly all pixels through the simple choice of observing only at the two extremal elevations, and observing the pixel when it is both rising and setting at each of these two elevations \footnote{The rising and setting aspect of this result is unsurprising as it is standard practice for ground-based surveys to observe fields when they are both rising and setting.}. We provide some details and explanation for why this is the case in appendix \ref{app_2ers}. 

In other words, for a given pixel, telescope latitude and minimum and maximum elevation bounds, this setup sets the optimal $h_k$ values that can be achieved in that pixel, for any detailed choice of scheduling; using any additional intermediate elevations will not improve the crossing-angle coverage. As a result, for each pixel, observatory latitude, and pair of bounding observing elevations, we can use the $|h_k|^2$ values calculated from the four crossing angles (two bounding elevations, rising and setting for each of them) as a bound on what can possibly be achieved in that pixel for any scan strategy composed of CESs, regardless of detailed scheduling choices. We call this approach for creating per-pixel lower bounds the two elevation rising and setting (2ERS) setup. We use this 2ERS setup in the rest of this work to create bounds on what is theoretically possible assuming CESs, independently of detailed scheduling choices. Note that 2ERS is not a scan strategy per se, as it is not a complete specification of the detailed scheduling choices.

The procedure used to arrive at the per pixel optimality of the 2ERS setup assumes that each pixel is scanned with at least two elevations. The case of only a single observing elevation is implicitly included in the 2ERS setup in the limit that the two elevations are arbitrarily close to each other, however for completeness, we explicitly consider the single elevation case in appendix \ref{sec_app}. Here we just note that even allowing for only a single elevation to be used, 2ERS is optimal when considering the whole visible field, in the sense that it minimises both the mean and the maximum of the distribution of $|h_k|^2$ values over the visible field.

For simplicity in the 2ERS setup we assume the elevations are equally weighted over the whole field. Generalising the analysis to include the possibility of weighting makes no difference to the utility of intermediate elevations. Whilst the two endpoints can be weighted differently, in practice doing so makes little difference unless one considers the extreme case, where it approaches the limiting case of observing with a single elevation. This is the case covered in appendix \ref{sec_app}.

\section{\lowercase{$h_k$} values from ground based surveys}
\label{sec_skyrot}
Having established that the 2ERS setup gives the optimal values of the $|h_k|^2$ parameters in each pixel given the bounding elevations at which observations are taken (for \textit{any} scan strategy composed of CESs with at least two observing elevations), we can now make maps of these optimal values, and investigate their declination and elevation dependence. We do this for $|h_k|^2$ with $k=1$ to $4$.

We work within the elevation range $30-65^\circ$, and we set the Latitude to $-22^\circ 56.396'$ corresponding to the Atacama Desert, a common site for CMB experiments. However, note that due to the symmetry of the situation, Northern Latitudes ($+\phi^\circ$) behave the same as Southern Latitudes ($-\phi^\circ$).

\subsection{Maps and declination dependence}
\begin{figure}
  \centering
  \includegraphics[width=\columnwidth]{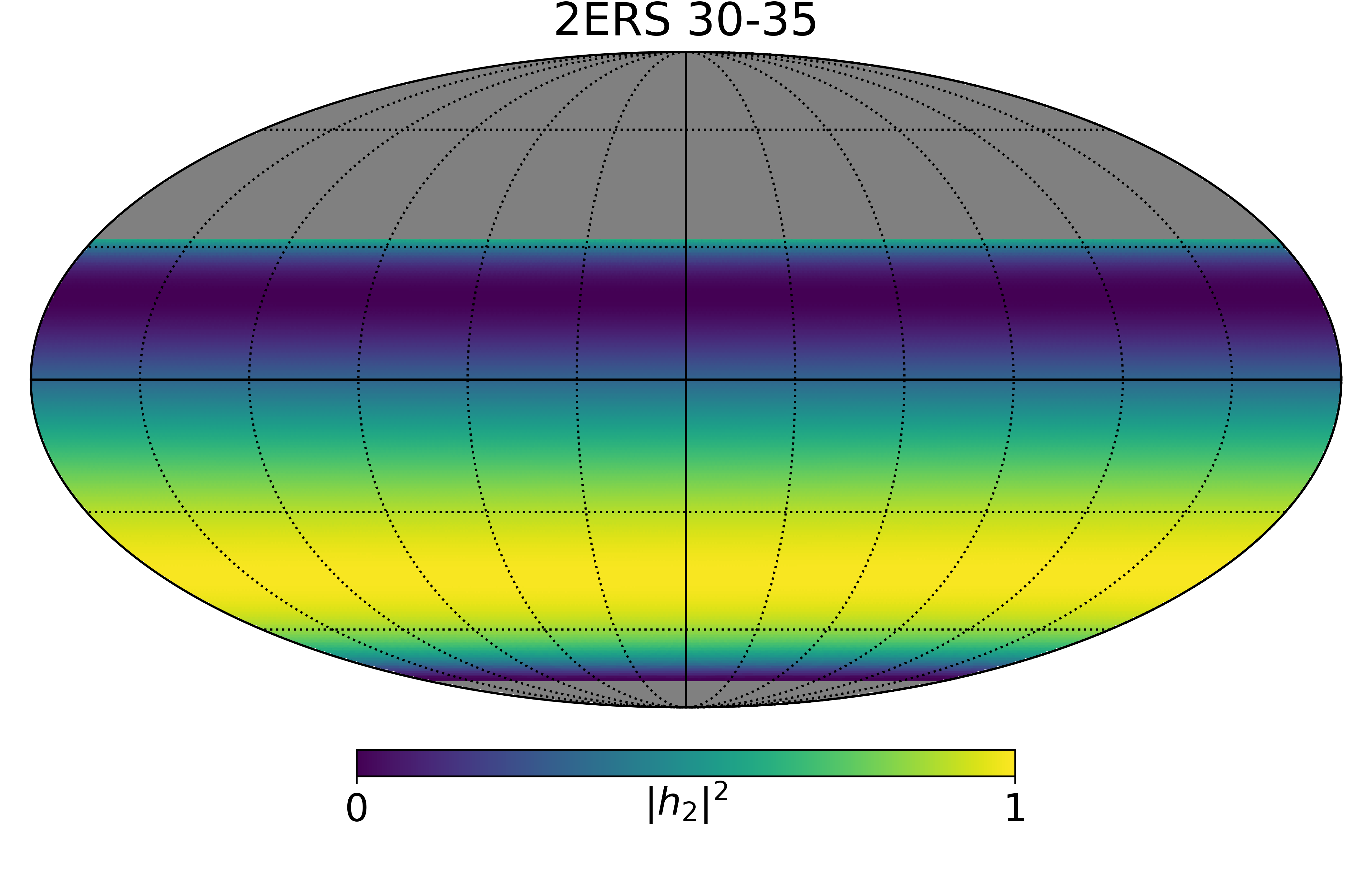}\\
    \includegraphics[width=\columnwidth]{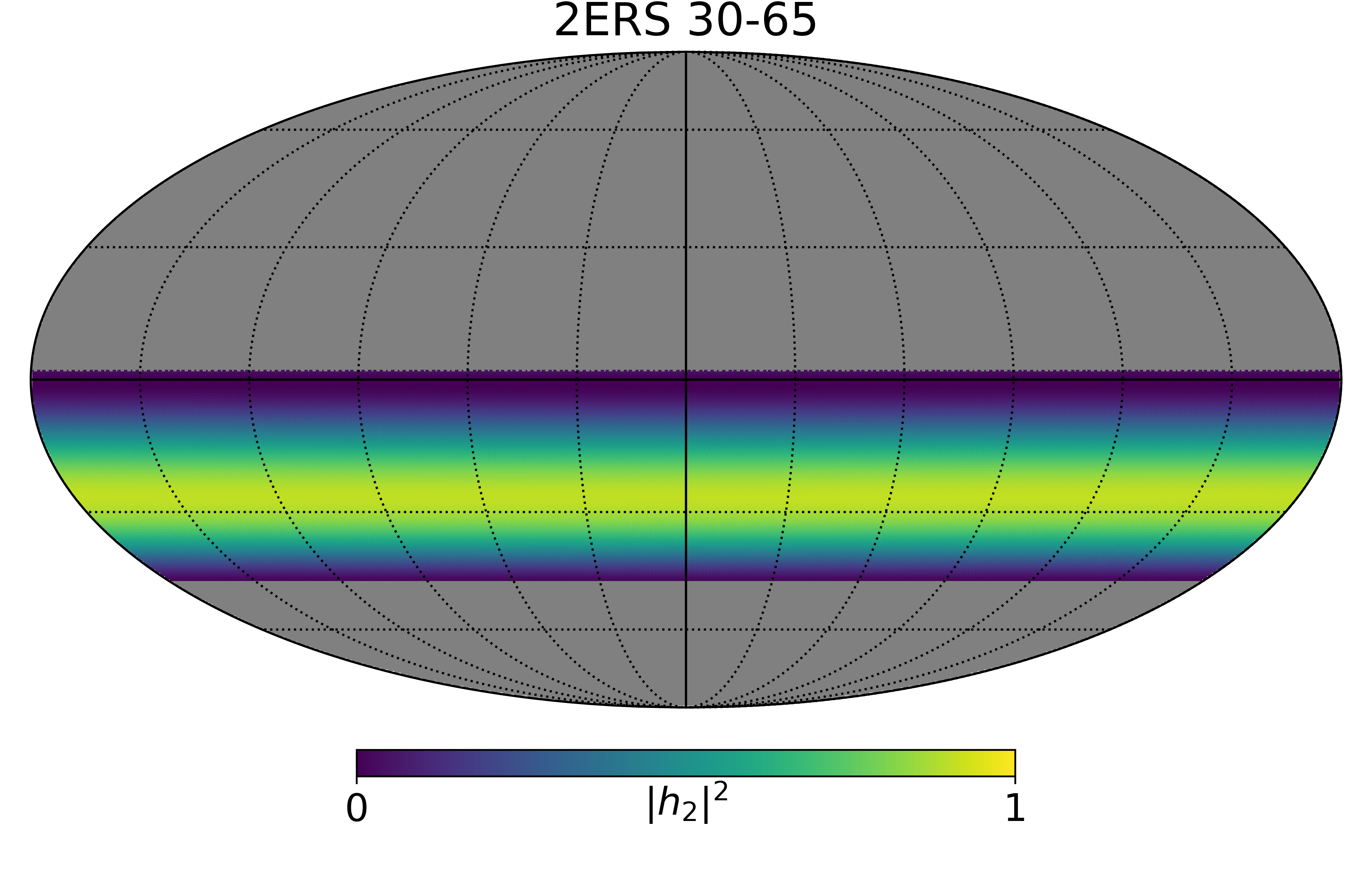}\\
   \includegraphics[width=\columnwidth]{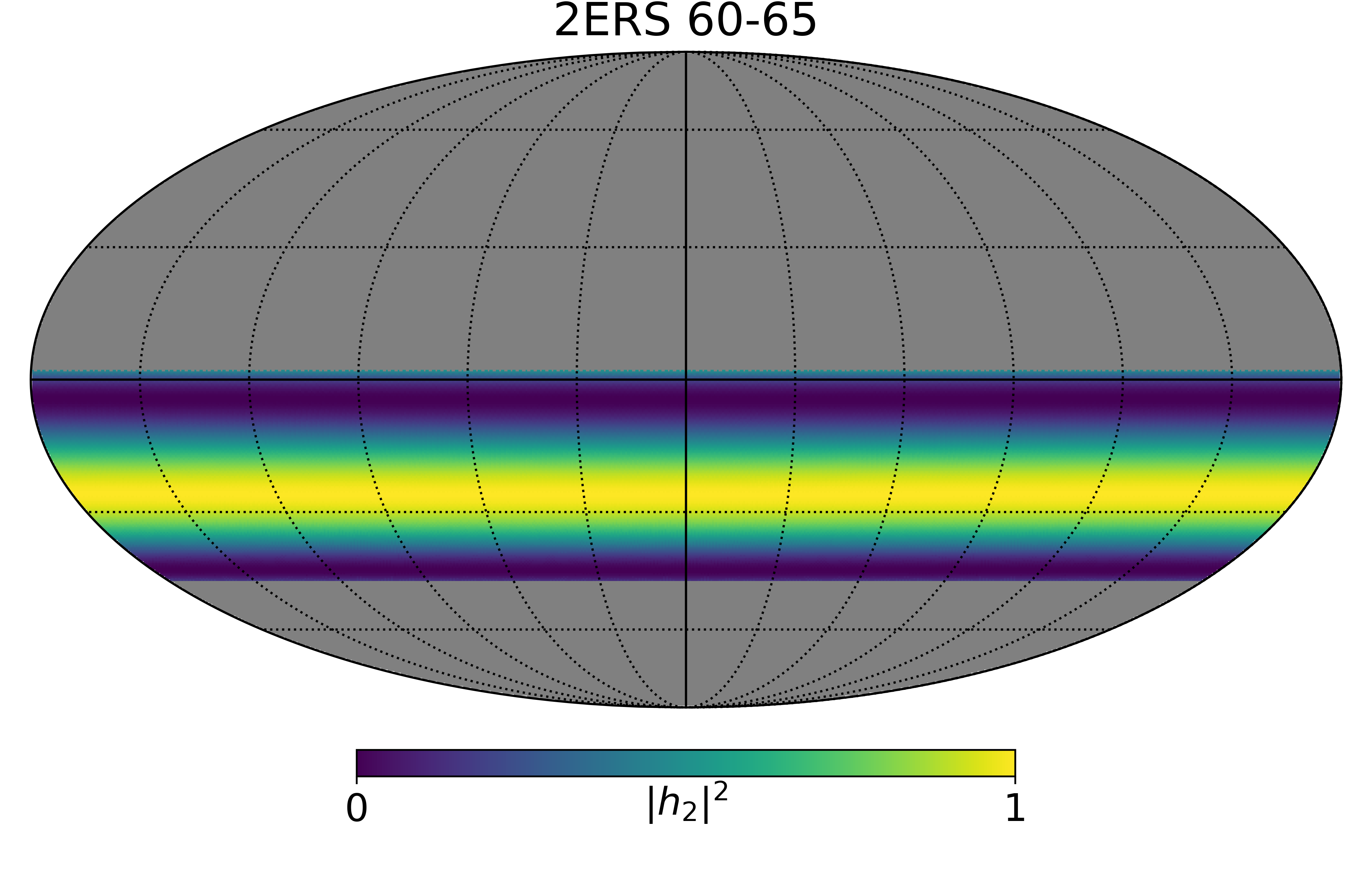}
\caption{$|h_2|^2$ maps for a telescope at Latitude $-22^\circ 56.396'$ for the 2ERS setup with elevation bounds (from top to bottom) $(30,35)$, $(30,65)$, and $(60,65)$. In all cases, there is a band of declinations within the visible area where the crossing angle coverage is at its worst. This band moves with the increase in visible area due to the upper elevation bound, and the worst values improve as the difference between the two elevation bounds increases. The maps are in equatorial coordinates.
}
\label{fig_solat}
\end{figure}

We show maps of $|h_2|^2$ in figure \ref{fig_solat}, computed using the 2ERS setup for three different pairs of elevations. There are several main features of these maps. The first is that the range of pixels seen by both elevations reduces as the upper elevation is increased, as expected. The second is that there is a clear pattern to the $h_2$ maps as a function of Dec, most notably a band of declinations within the visible area where the parameter is worst (largest). These features persist for all pairs of elevation bounds in the range $30-65^\circ$. Finally, the peak $|h_2|^2$ value for each map is lowest (i.e. best) for the map with the widest range of observing elevations: the peak values across the field are $0.99$, $0.91$ and $1.0$ for the elevation pairs $(30^\circ,35^\circ)$, $(30^\circ,65^\circ)$, and $(60^\circ,65^\circ)$ respectively. The peak values for various bounding elevations are shown in appendix \ref{app_peakvalues}.

A key consequence of these maps is that $|h_2|^2$ from any scan strategy composed of CESs is unlikely to be reasonably homogeneous over a large field. Moreover, since the 2ERS map shows the optimal value that can be achieved, making the field homogeneous would in practice simply amount to making the value of $|h_2|^2$ worse over most of the field. For any scan strategy composed of CESs, there are parts of the sky where $|h_2|^2$ will always be close to one and cannot be significantly reduced, in contrast with satellite scan strategies. We reiterate that, due to the correspondence with the cross-linking metric used in \cite{actmetric}, the achievable cross-linking is limited in the same way as these limitations on $|h_2|^2$.

\begin{figure}
  \centering
  \includegraphics[width=\columnwidth]{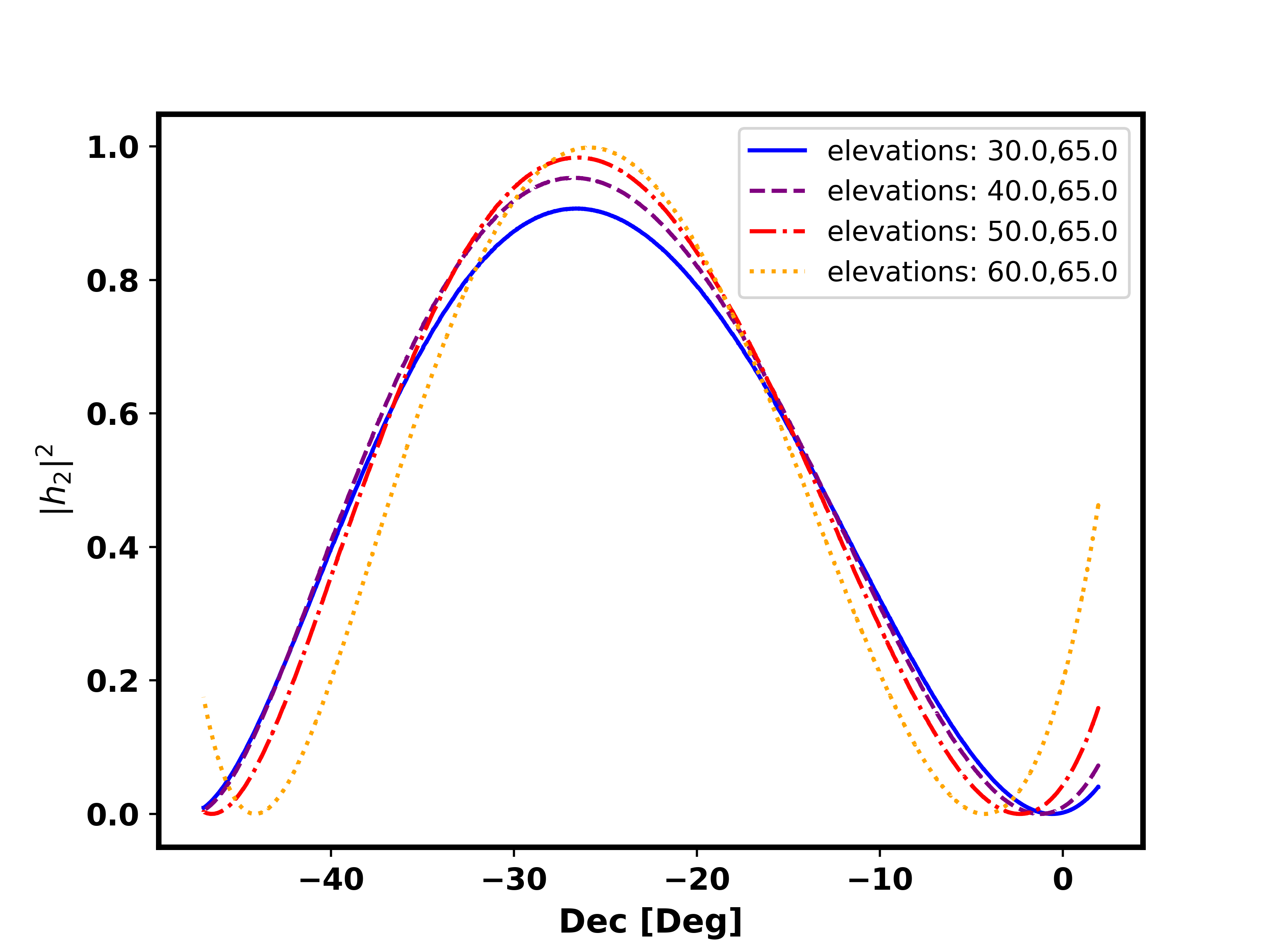}\\
   \includegraphics[width=\columnwidth]{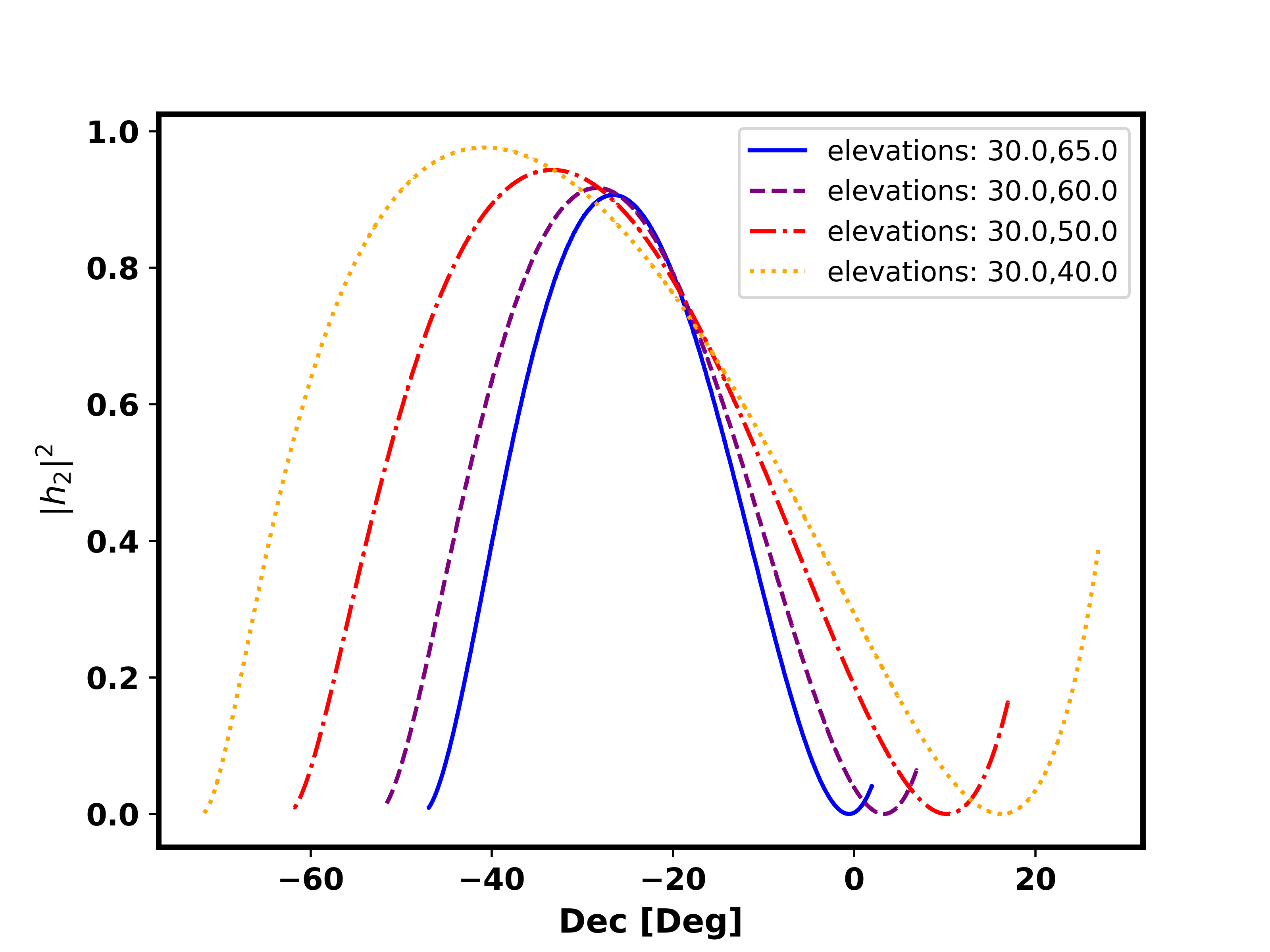}
\caption{$|h_2|^2$ as a function of declination as the elevation bounds are changed for the 2ERS setup and Latitude $-22^\circ 56.396'$. Top: the upper elevation is fixed to $65^\circ$ and the lower bound is varied as $30^\circ$ (blue, solid), $40^\circ$ (purple, dashed), $50^\circ$ (red, dot-dashed), $60^\circ$(oranged, dotted). Bottom: the lower elevation is fixed to $30^\circ$ and the upper bound is varied as $65^\circ$ (blue, solid), $60^\circ$ (purple, dashed), $50^\circ$ (red, dot-dashed), $40^\circ$(orange, dotted). In both cases, increasing the difference between the elevations reduces the peak. The lower plot also shows the decrease of visible area, and movement of the peak, as the upper elevation is increased.
}
\label{fig_1dsolat}
\end{figure}

We examine $|h_2|^2$ in more detail in figure \ref{fig_1dsolat}, where we have taken 1D slices through the maps at RA$=180^\circ$ (this choice is arbitrary and a different choice does not affect the results). The top plot shows the variation of $|h_2|^2$ with declination as the lower elevation bound is varied, and the bottom plot shows the variation of $|h_2|^2$ as the upper elevation bound is varied. The lower plot shows the decrease in visible area, and slight shift of the peak location, as the upper elevation increases. The decrease in the peak (worst) values with increasing difference between the elevation bounds is clear in both plots. This peak decrease is not large, meaning that it will not result in a significant reduction in the effect of instrumental systematics. However, the connection between $|h_2|^2$ and cross-linking means that this peak reduction may be important for any map-making methods that have problems as $|h_2|^2$ approaches unity. Precisely quantifying the required level of cross-linking for these methods is complicated. Since our focus is on instrumental systematics, we leave a detailed investigation of this to future work. However, we reiterate that the bounds calculated in this section are independent of detailed scheduling choices, so if a detailed study finds these bounds insufficient for a particular map making method, \textit{no} scan strategy composed of CESs can deliver sufficient cross-linking.

\begin{figure}
  \centering
    \includegraphics[width=\columnwidth]{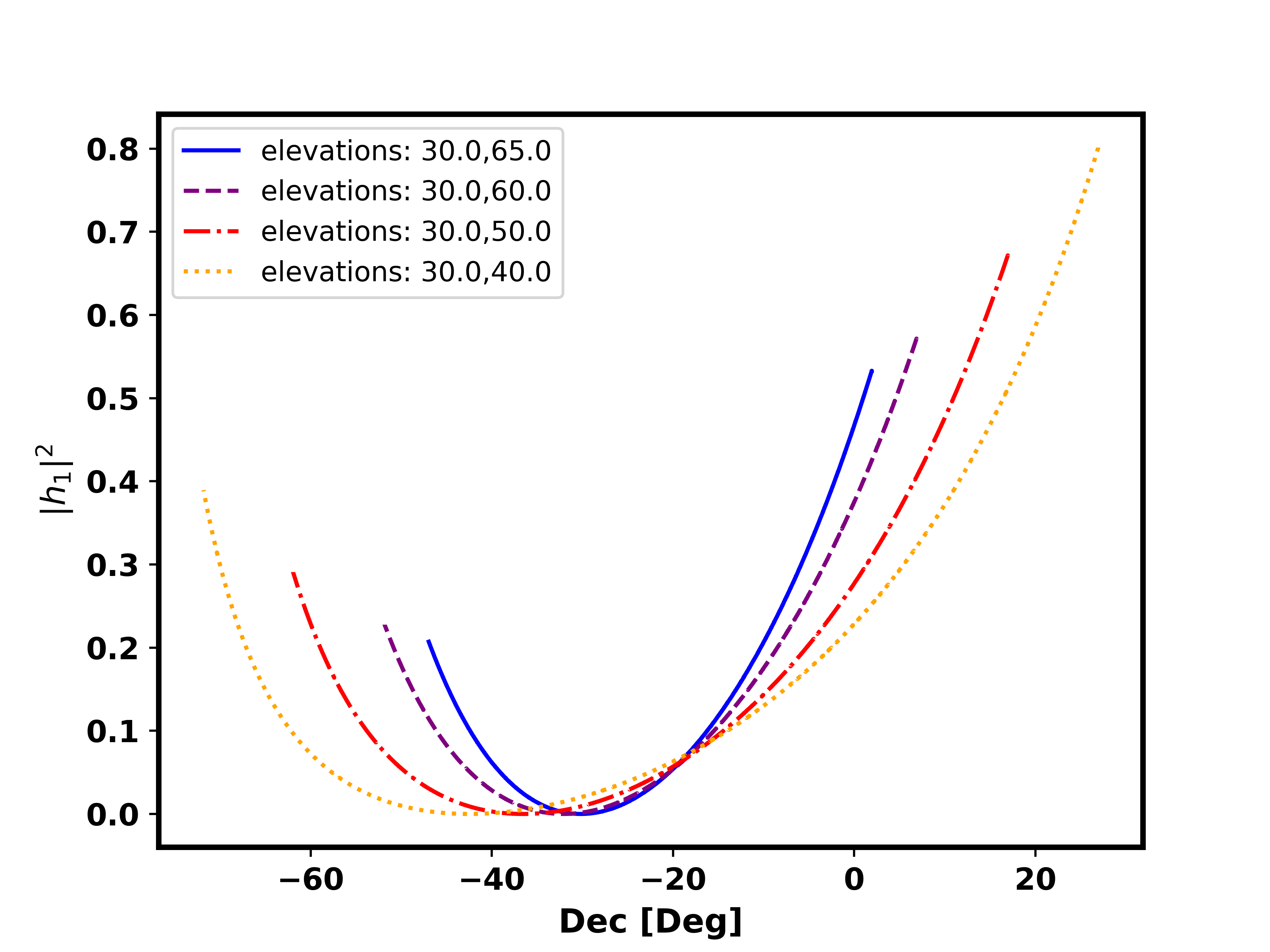}\\
   \includegraphics[width=\columnwidth]{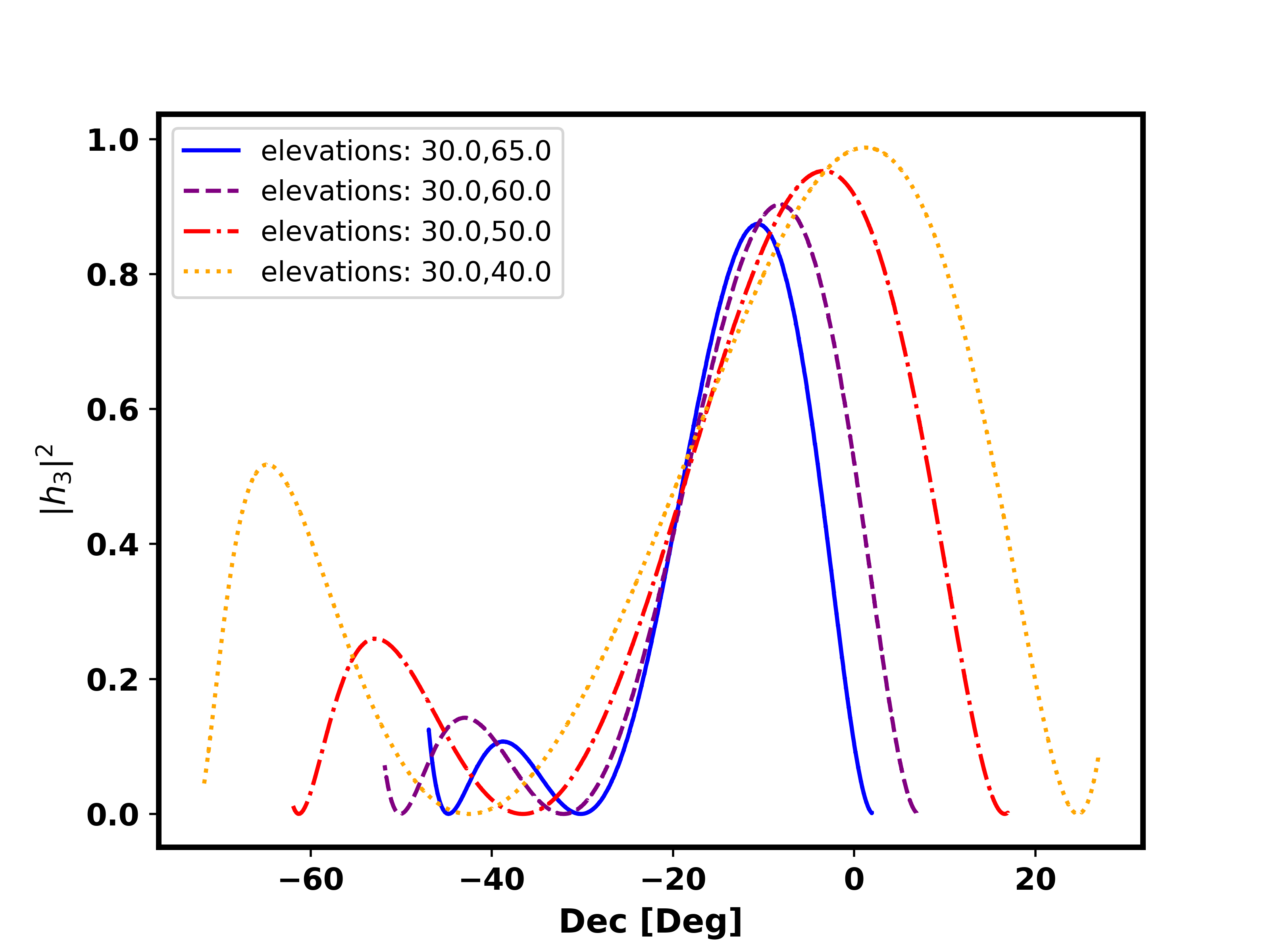}\\
   \includegraphics[width=\columnwidth]{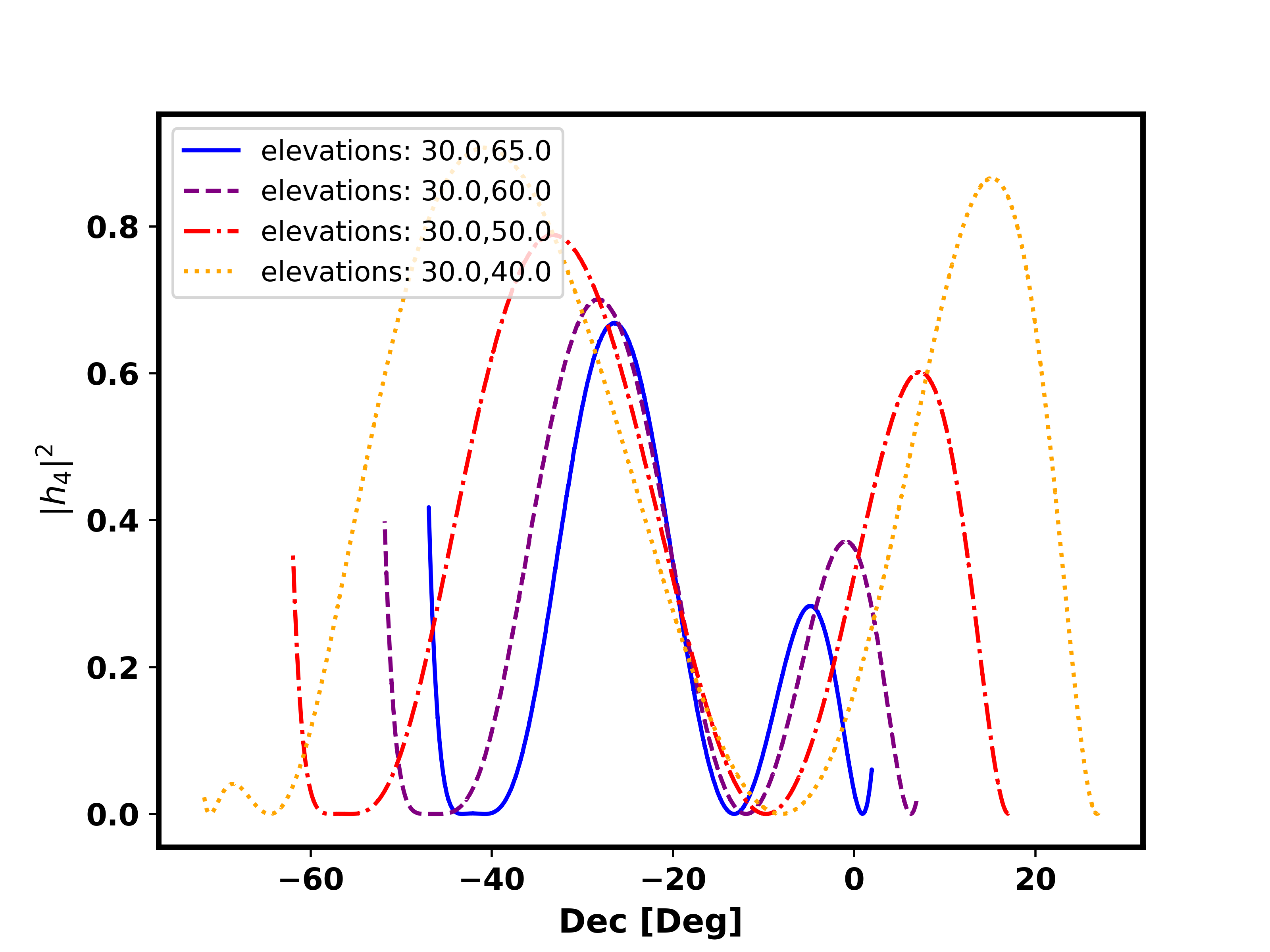}
\caption{$|h_1|^2$ (top), $|h_3|^2$ (middle) and $|h_4|^2$ (bottom) as a function of declination as the upper elevation bound is changed for the 2ERS setup and Latitude $-22^\circ 56.396'$. The lower elevation is fixed to $30^\circ$ and the upper bound is varied as $65^\circ$ (blue, solid), $60^\circ$ (purple, dashed), $50^\circ$ (red, dot-dashed), $40^\circ$(orange, dotted). As with $|h_2|^2$, a greater difference between the elevation bounds lowers the peaks, and increasing the upper elevation decreases the visible area and moves the position of the peak.
}
\label{fig_solat3}
\end{figure}

Figure \ref{fig_solat3} is the same as the lower panel of figure \ref{fig_1dsolat}, except that we are now plotting $|h_1|^2$, $|h_3|^2$ and $|h_4|^2$. Again, the 2ERS setup is used, with a lower elevation bound of $30^\circ$ and varying upper elevation bound, and we use the same latitude as before. Although the shape of each function is different to $|h_2|^2$, $|h_3|^2$ and $|h_4|^2$ also possess peak regions where the instrumental systematics cannot be significantly mitigated. Also similarly to $|h_2|^2$, the peak (worst) values are reduced when there is a greater difference between the elevation bounds. In addition, as for $|h_2|^2$, there is a small shift in the shape of the curve and the location of its peak as the upper elevation bound is changed.

The behaviour of $|h_1|^2$ is different. The best (minimum) values across the field are lower for a greater difference between the elevation bounds. As per the other parameters, $|h_1|^2$ has regions where the values are unavoidably poor, however for $|h_1|^2$ these occur for the declinations at the edges of the visible field and are reduced by using lower elevations. Lower elevations are also better for minimising $|h_1|^2$ when looking generally across the whole visible field.

The different shapes of the different curves mean that a declination range chosen to minimise one of these parameters, will not necessarily be minimal for the others.

We now briefly recap the results of this subsection. Using the 2ERS approach, we have shown that there is a strong declination dependence to the optimal values of these parameters in each pixel when CESs are used; this horizontal stripe structure will persist for any detailed scheduling choices. In particular, there are declinations for all four parameters for which the optimal values are poor, no matter which observing elevations one chooses (although for $|h_2|^2$, $|h_3|^2$ and $|h_4|^2$, a wider range of observing elevations slightly reduces these worst values, and for $|h_1|^2$ choosing lower elevations does the same). These poor values apply irrespective of detailed scheduling choices.

\subsection{$h_k$ tables for systematics}
The variation of the $|h_k|^2$ parameters with declination in the previous subsection is useful for understanding the possible structures that scan strategies can cause in maps due to the different systematics, the optimal values that can be achieved in each pixel, and how field selection can affect the contamination caused by these systematics. We now  consider the power spectrum contamination caused by the scan coupled systematics; this depends on the average of the respective $|h_k|^2$ over all of the observed pixels \citep{wallis}. We can use the 2ERS setup defined above to calculate the optimal values of these parameters averaged over a large observed field consisting of many pixels, and present these in tables. We assume the same elevation bounds are used in every pixel in the observed field.

We further validated the optimality of our results and the assumptions in the 2ERS setup by repeating the simulations for these tables whilst adding a third observing elevation. This third elevation was optimally chosen for each pixel to give a crossing angle midway between the two rising crossing angles from the elevation bounds. In all cases the average $|h_{k}|^2$ values across the fields were the same or worse than for the 2ERS case, although they typically do not degrade by much (up to 0.06).

Given a pair of common bounding elevations (and the observable field corresponding to the table), the value in the table is the lower bound on what can be achieved by any detailed scheduling choices. There are several uses for the tables we present. Firstly, for a given field, the values within the table corresponding to that field can be compared to determine the better elevations to use for observations, and how much of an effect it has to choose different observing elevations. The scan strategies from detailed scheduling choices can be compared to these values to determine how close they are to the limit of what can be achieved. Secondly, tables for different fields can be compared to assist in field selection for surveys that have freedom in their choice of which declinations to observe. Thirdly, the values in each table can be compared to $1$ to determine how much value there is in trying to optimise the detailed scheduling choices, and how close to an ideal scan can be reached for the field represented by that table. Finally, the values (and maps from earlier) can be used to forecast the best case and worst case scenarios for particular instrumental systematics, and thus the level of contamination that can be expected in future surveys, and the possible gain from reducing the instrumental imperfections.

In tables \ref{table_2ers_h1}-\ref{table_2ers_h4} we show the $|h_{k}|^2$ values (for $k=1$ to $4$)
calculated using the 2ERS setup at latitude $-22^\circ 56.396'$, and averaged over the entire observed field. In this case the observed field is the declination range $-46.96^\circ$ to $1.94^\circ$, which is the declination range visible to all elevation pairs in the tables. For completeness and for reference we include single elevation scans in the tables.

There are two key points from these tables. Firstly, the improvement compared to the worst possible scan ($|h_k|^2=1$) in even the most optimal cases is typically only a factor of a few (a factor of $2.13$ for $h_2$, $3.0$ for $h_3$ and $4.35$ for $h_4$). The improvement is better for $h_1$, where it can be up to a factor of $12.5$ compared to $|h_1|^2=1$. These improvement factors are worse for less optimal elevation choices. Given that the values in these tables are approximate bounds on what can be achieved for \textit{any} choice of detailed scheduling, and may not be reachable in practice, these values (even for $h_1$) are very poor compared to satellite scans, which can reduce these parameters by $2-3$ orders of magnitude.

Secondly, the variation amongst all of the entries within a particular table is relatively small, less than a factor of 4 in all cases (a factor of $3.75$ for $h_1$, $1.53$ for $h_2$, $1.67$ for $h_3$ and $2.26$ for $h_4$). This is relatively small given the wide range of elevation values being considered, and when compared to the variation in values amongst satellite surveys.

Generally, $|h_1|^2$ is the easiest to optimise and can be optimised the most. However, given that the 2ERS setup is a bound that ignores many of the practicalities of detailed scan strategy choices and scheduling, and the relatively modest reductions that are achieved in $h_k$ even for this ideal case, from these results we expect that it is not advisable for CMB experiments using CESs to prioritise optimising any of these parameters.

For $|h_2|^2$, $|h_3|^2$ and $|h_4|^2$ these tables show a similar general trend as the previous section that increasing the difference between the elevation bounds improves the quality of the survey in terms of its inherent mitigation of systematics. For $|h_1|^2$, the trend is as expected from the earlier maps and plots: lower elevations are superior.

The diagonal entries of each table show that restricting to a single elevation is never better (averaged over the whole visible area) than the best 2ERS setup.

There are some possible complications here that could affect these results. For a field where the edges cannot be seen at the observing elevations used for the central part of the field, the central part of the field will behave as described above. The edge regions can then be treated as one or more subsets in declination, with a maximum elevation that is reduced in steps from that used in the central field. Each step between the central region and the edge of the full field will behave similarly to the central region, just with a different set of elevation bounds. I.e. one can in effect stitch together several regions. For simplicity here we have focused on exploring generally how the individual regions behave. A related point is that we have not applied any masks; a mask that systematically and significantly affects some declinations more than others could change the results in these tables, however most masks and apodisation schemes contain a substantial central region that is treated uniformly, so it would require a pathological case to substantially change the results. Due to the assumption that the same elevation bounds are used in every pixel, the numbers in the tables will not be strict lower bounds; in principle, small improvements could be made by making bespoke per pixel elevation choices. We have checked numerically that such per pixel choices do not have a significant effect on the optimal values compared to the best case where the pixels use the same elevations, and this effect does not change our conclusions. Generally, we note that the caveats to the optimality discussed here are primarily a formality: the bounds we present here are close to optimal and it is unlikely that these bounds can be achieved in practice. Moreover, given the other constraints that exist on scan strategies, attempting to design a scan strategy that achieves the small gains that could be achieved in principle from exploiting these caveats is likely to result in something that is unrealistic and poor in other respects. As a result, these caveats do not affect the conclusions that will be drawn in this work.

The approach and codes used to make these tables are general, and can be used to create tables for any observatory targeting any declination range. For example, we have re-run our analysis restricting to different subsets of the full declination range in the tables presented here. The results are broadly similar to the results presented here, with each $|h_k|^2$ being better in one half of the full declination range and worse in the other half, as expected from figures \ref{fig_1dsolat} and \ref{fig_solat3}. The variation of the values between different subsets of declination is typically larger than the variation due to differing elevations for a fixed declination range.

Since the chosen latitude for our results is the Atacama Desert, the bounds in this section represent the best possible (smallest) $|h_k|^2$ (averaged over the visible sky area) that can be achieved from sky rotation by experiments in the Atacama Desert (such as ACT or Simons Observatory) using CESs. In the same fashion, bounds can be constructed for any ground-based observatory at a different latitude using our approach.

\begin{table}
 \caption{$|h_1|^2$ values for different elevation bounds using 2ERS setup, averaged over the entire sky between declinations $-46.96^\circ$ and $1.94^\circ$.}
\centering
\begin{mytabular}[1.8]{|c|cccccccc|} 
 \hline 
 & $30^\circ$ & $35^\circ$ & $40^\circ$ & $45^\circ$ & $50^\circ$ & $55^\circ$ & $60^\circ$ & $65^\circ$  \\
\hline 
$30^\circ$ & $0.08$ & $0.08$ & $0.08$ & $0.08$ & $0.09$ & $0.10$ & $0.12$ & $0.15$\\
$35^\circ$ &-& $0.08$ & $0.08$ & $0.09$ & $0.09$ & $0.10$ & $0.12$ & $0.15$\\
$40^\circ$ &-&-& $0.08$ & $0.09$ & $0.10$ & $0.11$ & $0.13$ & $0.16$\\
$45^\circ$ &-&-&-& $0.09$ & $0.10$ & $0.12$ & $0.14$ & $0.17$\\
$50^\circ$ &-&-&-&-& $0.11$ & $0.13$ & $0.15$ & $0.19$\\
$55^\circ$ &-&-&-&-&-& $0.15$ & $0.17$ & $0.21$\\
$60^\circ$ &-&-&-&-&-&-& $0.20$ & $0.25$\\
$65^\circ$ &-&-&-&-&-&-& - & $0.30$\\
 \hline 
 \end{mytabular} 
 \label{table_2ers_h1}
\end{table}

\begin{table}
 \caption{$|h_2|^2$ values for different elevation bounds using 2ERS setup, averaged over the entire sky between declinations $-46.96^\circ$ and $1.94^\circ$.}
\centering
\begin{mytabular}[1.8]{|c|cccccccc|} 
 \hline 
 & $30^\circ$ & $35^\circ$ & $40^\circ$ & $45^\circ$ & $50^\circ$ & $55^\circ$ & $60^\circ$ & $65^\circ$  \\
 \hline 
$30^\circ$ & $0.71$ & $0.72$ & $0.71$ & $0.70$ & $0.68$ & $0.63$ & $0.57$ & $0.48$\\
$35^\circ$ &-& $0.72$ & $0.72$ & $0.71$ & $0.68$ & $0.64$ & $0.58$ & $0.49$\\
$40^\circ$ &-&-& $0.72$ & $0.71$ & $0.68$ & $0.64$ & $0.58$ & $0.50$\\
$45^\circ$ &-&-&-& $0.70$ & $0.67$ & $0.63$ & $0.57$ & $0.50$\\
$50^\circ$ &-&-&-&-& $0.65$ & $0.61$ & $0.56$ & $0.49$\\
$55^\circ$ &-&-&-&-&-& $0.58$ & $0.54$ & $0.48$\\
$60^\circ$ &-&-&-&-&-&-& $0.50$ & $0.47$\\
$65^\circ$ &-&-&-&-&-&-& - & $0.47$\\
 \hline 
 \end{mytabular} 
   \label{table_2ers_h2}
\end{table}

\begin{table}
 \caption{$|h_3|^2$ values for different elevation bounds using 2ERS setup, averaged over the entire sky between declinations $-46.96^\circ$ and $1.94^\circ$.}
\centering
\begin{mytabular}[1.8]{|c|cccccccc|} 
 \hline 
 & $30^\circ$ & $35^\circ$ & $40^\circ$ & $45^\circ$ & $50^\circ$ & $55^\circ$ & $60^\circ$ & $65^\circ$  \\
 \hline 
$30^\circ$ & $0.52$ & $0.49$ & $0.48$ & $0.46$ & $0.46$ & $0.44$ & $0.41$ & $0.33$\\
$35^\circ$ &-& $0.48$ & $0.47$ & $0.46$ & $0.46$ & $0.45$ & $0.42$ & $0.34$\\
$40^\circ$ &-&-& $0.46$ & $0.47$ & $0.47$ & $0.47$ & $0.44$ & $0.36$\\
$45^\circ$ &-&-&-& $0.48$ & $0.49$ & $0.50$ & $0.47$ & $0.39$\\
$50^\circ$ &-&-&-&-& $0.51$ & $0.53$ & $0.51$ & $0.42$\\
$55^\circ$ &-&-&-&-&-& $0.55$ & $0.54$ & $0.46$\\
$60^\circ$ &-&-&-&-&-&-& $0.54$ & $0.48$\\
$65^\circ$ &-&-&-&-&-&-& - & $0.48$\\
 \hline 
 \end{mytabular}
  \label{table_2ers_h3}
\end{table}

\begin{table}
 \caption{$|h_4|^2$ values for different elevation bounds using 2ERS setup, averaged over the entire sky between declinations $-46.96^\circ$ and $1.94^\circ$.}
\centering
\begin{mytabular}[1.8]{|c|cccccccc|} 
 \hline 
 & $30^\circ$ & $35^\circ$ & $40^\circ$ & $45^\circ$ & $50^\circ$ & $55^\circ$ & $60^\circ$ & $65^\circ$  \\
 \hline 
$30^\circ$ & $0.34$ & $0.38$ & $0.40$ & $0.39$ & $0.37$ & $0.34$ & $0.29$ & $0.23$\\
$35^\circ$ &-& $0.42$ & $0.44$ & $0.44$ & $0.42$ & $0.38$ & $0.33$ & $0.26$\\
$40^\circ$ &-&-& $0.46$ & $0.47$ & $0.45$ & $0.42$ & $0.37$ & $0.29$\\
$45^\circ$ &-&-&-& $0.47$ & $0.46$ & $0.44$ & $0.41$ & $0.32$\\
$50^\circ$ &-&-&-&-& $0.47$ & $0.46$ & $0.44$ & $0.35$\\
$55^\circ$ &-&-&-&-&-& $0.48$ & $0.49$ & $0.40$\\
$60^\circ$ &-&-&-&-&-&-& $0.52$ & $0.47$\\
$65^\circ$ &-&-&-&-&-&-& -& $0.49$\\
 \hline 
 \end{mytabular} 
  \label{table_2ers_h4}
\end{table}

\subsection{Discussion}
From the previous two subsections, we can now answer some of the questions posed earlier in the paper. The key quantities determining the $h_k$ values in each pixel are the declination and the number of different observing elevations used\footnote{And any detailed scheduling choices that result in non-uniform distribution of the hits within a pixel between the different elevations and between rising and setting}. Typical values within a pixel for $|h_k|^2$ for ground-based surveys using CESs are between $0.1$ and $1$.

When considering the average of $|h_k|^2$ over the entire observable field, the possible improvement due to the scan strategy is not significant. Moreover, the variation with scan strategy of $|h_k|^2$ averaged over an observable field is relatively slight: there is a fundamental limit to how well these quantities can be optimised, and this limit is orders of magnitude worse than the values that can be achieved by satellite scans. In particular an ideal scan cannot be approached for ground-based surveys, so care must be taken if this is assumed whilst forecasting or modelling ground based surveys.

For a particular instrumental systematic, the values in the tables can be used to evaluate how much further optimisation might be possible for the relevant $h_k$ for a given detailed choice of scheduling. The bounds in tables \ref{table_2ers_h1}-\ref{table_2ers_h4}, and the maps in figure \ref{fig_solat} (and the equivalent maps for $h_1$, $h_3$ and $h_4$ with a mask applied if required) can be used in combination with the expressions in \cite{wallis} \& \cite{formalism} to predict the magnitude of the distortion due to scan-coupled systematics on both CMB power spectra and CMB maps.

The relatively narrow range of possible values, and closeness to 1, means that the effect of these systematics can be calculated and forecasted relatively accurately \textit{without} requiring any detailed information about how the instrument in question will scan the sky. This ability is particularly useful earlier in the forecasting and design process to get approximate values and best case scenarios, that can be used to inform field selection and the other design aspects of the surveys. These results can also be used to calculate the threshold or tolerance at which the various instrumental systematics will cause leakage that will affect the science goals.

For example, a differential pointing systematic causes a temperature to polarisation leakage modulated by $h_1$ and $h_3$, with a magnitude $\rho$ (for certain focal plane setups) \citep{formalism, wallis},
\begin{equation}
\Delta C^B_\ell=\frac{1}{8}\ell^2C^T_\ell \rho^2 \left(\langle|h_1|^2\rangle+\langle|h_3|^2\rangle \right) \text{,}
\end{equation}
where the angle brackets denote the average value over the observed pixels. To determine the magnitude of $\rho$ at which the contamination will reach a certain level thus requires knowledge of the scan strategy. One can assume the worst case for the scan strategy (i.e. $|h_1|^2+|h_3|^2=2$), but for satellite scan strategies this would give the wrong requirements on $\rho^2$ by 3 orders of magnitude. From the tables we have shown here, a ``best case'' scenario can now be computed for ground-based experiments. For example, it can be seen that over the whole field, the absolute best case scenario for the scan strategy ($|h_1|^2+|h_3|^2=0.48$) only reduces the requirement on $\rho^2$ by a factor of $2.1$, showing a stark difference to the satellite case. This range of best to worst improvement of $2.1$ to $1$ can be used to forecast the effect of this systematic for Atacama-based CMB-S4 surveys (or any other Atacama-based surveys), without requiring any detailed scheduling choices.

\subsection{Combined sky rotation and  boresight rotation}
\label{sec_instrotate}
Many CMB experiments use boresight rotation to mitigate systematics and provide null tests (see e.g. \cite{bicepsyst}). We can use the 2ERS setup to get an estimate of how sky rotation interacts with boresight rotation, and estimate the relative efficacy of the two approaches independent of detailed scheduling choices. We consider the case of a single boresight rotation angle: whilst many (typically smaller aperture) CMB experiments include multiple rotation angles, there are cases where only a single angle is possible, such as the SO large aperture telescope \citep{pbr,1808.05131}. In principle we could extend this to many rotation angles, for example the set of eight angles separated by $45^\circ$ as used by BICEP \citep{bicep2015}. We will see that we only need to consider a single angle to show the superiority of boresight rotation over sky rotation, irrespective of detailed scheduling choices. Note that in this section $h_2$ no longer relates to the cross-linking, because boresight rotation changes the orientation of the focal plane with respect to the sky without changing the path of the instrument across the sky.

We consider the case where the instrument has two values for the boresight rotation, taken to be $0^\circ$ and the boresight rotation value. For each crossing angle in the 2ERS setup, we now include a pair of angles, where the second is adjusted according to the rotated value. In some cases, the results follow trivially from the symmetry of the $\cos$ and $\sin$ functions, as taken advantage of in e.g. \cite{pbr}. For example, $180^\circ$ rotation has no effect on $h_2$, but reduces $h_1$ to zero. We consider three angles: $180^\circ$, $90^\circ$ and $45^\circ$, and present their effects on the different metrics in table \ref{table_pbr}. Note that the ranges in this table are showing the approximate range of improvement achieved over the 28 elevation combinations considered in the tables earlier, rather than showing the result for a single choice of bounding elevations. The key result here is that even the inclusion of a single boresight rotation angle can result in a strong improvement on the mitigation of scan-coupled systematics compared to solely using sky rotation. As with our earlier results, this is independent of the detailed scheduling choices, showing that the gain to $|h_k|^2$ from detailed optimisation of the scan strategy is of limited utility.

\begin{table}
 \caption{Effects on $h_k$ for different choices of boresight rotation angle using 2ERS setup.}
\centering
\begin{mytabular}[1.8]{|c|c|c|c|} 
 \hline 
 & $45^\circ$ & $90^\circ$ & $180^\circ$ \\
\hline 
$h_1$ & $1.5-2\times$ worse &  $1.5-2\times$ worse & reduced to zero \\
$h_2$ & $2-3\times$ better & reduced to zero & no effect \\
$h_3$ & $5-7\times$ better & $2\times$ better & reduced to zero \\
$h_4$ & reduced to zero & no effect & no effect \\
 \hline 
 \end{mytabular} 
 \label{table_pbr}
\end{table}

\section{Conclusions}
\label{sec_conc}
We have investigated the consequences of scan strategies composed of CESs for scan-coupled systematics. We used a method that does not require detailed scheduling choices or simulations, making the conclusions very general. This method allows us to answer the questions posed in the introduction. The key results and conclusions are
 \begin{enumerate}
  \item We have created a setup that we call 2ERS, which we have used to calculate per pixel lower bounds on the parameters that couple instrumental systematics to the scan strategy. We have shown that CESs put strong constraints on the possible values of these parameters that can be achieved, independently of detailed scheduling choices.
  \item Typical per pixel values of $|h_k|^2$ for $k=1$ to $4$ are between $0.1$ and $1$, regardless of detailed scheduling choices. There is a fundamental declination dependence of these scan coupling parameters caused by the use of CESs. This horizontal stripe structure will persist in the maps for any realistic choice of detailed scheduling. In particular, for all four parameters, there are declination ranges with unavoidably poor values.
 \item The optimal (lowest) values of these parameters are driven by the choices of observing elevations and the declination of the targeted field, however these optimal values cannot approach an ideal scan, unlike the case of satellite surveys. The variation with elevation of the optimal achievable parameters is relatively small when averaged across the visible field, less than a factor of $4$ for all parameters, and less than a factor of $2$ for $|h_2|^2$ and $|h_3|^2$. In addition, the improvement compared to the worst possible case for $|h_2|^2$, $|h_3|^2$ and $|h_4|^2$ is at best a factor of $4.35$; $|h_1|^2$ can be improved by up to a factor of $12.5$.
 \item Combined, these results suggest that attempting to strongly optimise detailed scheduling choices to minimise these parameters is not advisable, particularly given the other constraints on realistic scan strategies. We have also shown that inclusion of even a single boresight rotation angle can result in a greater reduction in the contamination than any scan strategy built from CESs. 
 \item The maps and tables constructed using our approach in section \ref{sec_skyrot} can be used to calculate and forecast the approximate level of contamination of these systematics at both the map and power spectrum levels for \textit{any} CMB experiment based in the Atacama Desert, without requiring knowledge of detailed scan strategy choices. These values provide a guide to the relative importance of different systematics. These values can also be used to assess how far from optimal a particular detailed choice of scan strategy and scheduling is. Whilst we have focused on the Atacama Desert as the most important example, the codes and approach in this work can be easily applied to create such maps and tables for an observatory at any latitude.
 \item Generally, $|h_2|^2$, $|h_3|^2$ and $|h_4|^2$ are improved at a pixel or field level by considering more widely spaced elevations. However, $|h_1|^2$ behaves quite differently: it prefers lower elevations more than widely spaced elevations, and it can be optimised further than the other 3 parameters.
 \item We also note that the standard metric for cross-linking \citep{actmetric} is closely related to one of the parameters used in this work ($|h_2|^2$). As such, a corollary of our results is that there are fundamental limits to the cross-linking that can be achieved using CESs, regardless of any detailed scheduling choices. This also means that if homogeneous cross-linking is required, this can only be achieved by making the better areas worse, and not by improving the regions where cross-linking is bounded to be poor.
\end{enumerate}
We expect these results to be useful for understanding the limits of CESs for upcoming CMB surveys, particularly SO and CMB-S4 \citep{cmbs4}, and also for allowing forecasts and simulations for different systematics to be made, all without requiring detailed choices of scheduling and expensive simulations. In addition, these results apply to any non-CMB ground based surveys that perform CESs, where the spins of different quantities are important, including intensity mapping.

\acknowledgements
DBT acknowledges support from Science and Technology Facilities Council (STFC) grants ST/P000649/1, ST/T000414/1 and ST/T000341/1. NM is supported by an STFC studentship. MLB acknowledges support from STFC grant ST/T007222/1. The codes used to generate these results are available from the authors on request.

\appendix
\section{A. Justification and explanation of 2ERS optimality}
\label{app_2ers}
In the main text we comment that for a given range of observing elevations, as long as at least two observing elevations are used, the $|h_n|^2$ parameters are minimised in the majority of pixels by only observing at the two bounding elevations, and doing this for both rising and setting observations. In this appendix we explain why this is the case.

\subsection{Rising and setting}
The generic expression for $|h_n|^2$ in a pixel is given by
\begin{equation}
|h_n|^2_\text{rise}=\frac{1}{N^2}\left(\sum^N_i \cos(n\psi_i)\right)^2+\frac{1}{N^2}\left(\sum^N_i \sin(n\psi_i)\right)^2\text{.}
\end{equation}
Let us assume without loss of generality that the set of $N$ crossing angles in this sum are all rising. Given the relationship between rising and setting crossing angles (a sign change) and the fact that $\sin$ is an odd function and $\cos$ is an even function, we can consider how this expression changes when a second set of $N$ crossing angles are included, consisting of the setting counterparts of the initial set of angles,
\begin{eqnarray}
&&|h_n|^2=\frac{1}{4N^2}\left(\sum^{2N}_i \cos(n\psi_i)\right)^2+\frac{1}{4N^2}\left(\sum^{2N}_i \sin(n\psi_i)\right)^2\\
&&=\frac{1}{4N^2}\left(\sum^{N}_i \cos(n\psi_i)+\sum^{N}_i \cos(-n\psi_i)\right)^2+\frac{1}{4N^2}\left(\sum^{N}_i \sin(n\psi_i)+\sum^{N}_i \sin(-n\psi_i)\right)^2\\
&&=\frac{1}{4N^2}\left(\sum^{N}_i \cos(n\psi_i)^2+\sum^{N}_i \cos(n\psi_i)\right)^2=\frac{1}{N^2}\left(\sum^N_i \cos(n\psi_i)\right)^2\leq|h_n|^2_\text{rise}
\end{eqnarray}
Therefore it will never be advantageous for the $|h_n|^2$ parameters to not include the setting counterpart to any rising observations, and vice versa.

\subsection{Systematically spaced elevations}
Having established that we will always work with rising and setting crossing angle pairs, the next thing to establish is why additional elevations between the endpoints do not improve the $|h_n|^2$ parameters. We start with the expression for combined rising and setting angles, and we place equally spaced crossing angles between the endpoints (the endpoints are denoted $\psi_A$ and $\psi_B$)

\begin{equation}
|h_n|^2
=\frac{1}{N^2}\left(\sum^N_i \cos(n\psi_i)\right)^2=\frac{1}{N^2}\left(\sum^N_i \cos\left(n\left[\psi_A+\frac{(i-1)}{(N-1)}\left(\psi_B-\psi_A)\right)\right]\right)\right)^2
\end{equation}

We can now use the result that 
\begin{equation}
\sum^{N-1}_{i=0} \cos\left(a+ib \right)=\cos\left(a+\frac{(N-1)}{2}b \right)\frac{\sin\left(\frac{Nb}{2} \right)}{\sin\left(\frac{b}{2} \right)}\text{,}
\end{equation}
so for equally spaced angles we can express $|h_n|^2$ as a function of $N$ as
\begin{equation}
|h_n|^2=\frac{1}{N^2}\left(\sum^N_i \cos(n\psi_i)\right)^2\propto\left(\frac{1}{N}\frac{\sin\left(n \frac{N}{N-1}\frac{\psi_B-\psi_A}{2} \right)}{\sin\left(n \frac{1}{N-1}\frac{\psi_B-\psi_A}{2} \right)} \right)^2 \text{.}
\end{equation}
For $N\geq2$, this function is monotonically increasing with $N$, provided that $\psi_A-\psi_B<\frac{203.1}{n}$. This condition is satisfied for all pixels for $n=1,2$ and is satisfied for the overwhelming majority of pixels for $n=3,4$, with the exception of a small range of declinations at the edges of the visible field. As a result, the non-optimally of 2ERS for our results happens only for small declination ranges at the edge of the visible field, and even for most of these declinations 2ERS is close to optimal. Given the large sky areas targeted by upcoming CMB experiments, this will not significantly affect the results in the main text.

\clearpage
\section{B. Single elevation scans}
\label{sec_app}
The optimality of the 2ERS setup as discussed in appendix \ref{app_2ers} assumes at least two elevations are used. The case of a single elevation is implicitly included in the limit that the two bounding elevations are equal, or the limit when one of the bounding elevations is weighted significantly more than the other. For completeness, in this appendix we explicitly consider the case of a single elevation and see in which circumstances, and by how much, this could generate a more optimal bound than the 2ERS setup. Note the optimality of the 2ERS setup in the sense that intermediate elevations do not improve the $|h_k|^2$ parameters is unaffected by the discussion in this appendix.

Note there are clear limitations to scans with a single elevation: a maximum of 2 crossing angles within each pixel is achievable, and it limits the amount of time that the target field can be scanned. As such, it is only realistic for surveys targeting large patches of the sky, and we expect these to cover a larger declination range than the band in which a single elevation is the best choice. Of course one could observe a large field by observing different strips in declination with their respective optimal single elevation, but this would require much more complicated scheduling and will not result in significant reductions in $|h_k|^2$ compared to a 2ERS setup. As a result, the cases we will discuss below where there are possible small improvements over the case with two distinct elevations are not necessarily realistic.

In general, 2ERS with two distinct bounding elevations has the best average over the whole range of visible declinations, even when allowing for single elevation scans (see tables \ref{table_2ers_h1}-\ref{table_2ers_h4}). 

In addition, the peak (worst) values over the entire field will always be worse for the case of a single elevation compared to 2ERS with two distinct bounding elevations.

However, a single elevation can deliver smaller values of $|h_k|^2$ than 2ERS in specific pixels, or small declination ranges. This situation occurs when the targeted declination range (dependent on the latitude of the telescope) is carefully chosen and small enough; in this case the optimal $|h_k|^2$ values come from observing at only one of the two limiting elevations, and adding any other elevations will degrade the quality of the crossing-angle coverage. Consequently, if using two observing elevations, the bounds from the 2ERS setup will not quite be strict bounds. However, in practice, the average, peak, and per pixel values from considering a two distinct elevations are close to those from a careful choice of single elevation. As such, none of the conclusions in the main text are altered by considering this case.

\begin{figure}
  \centering
    \includegraphics[width=3.5in]{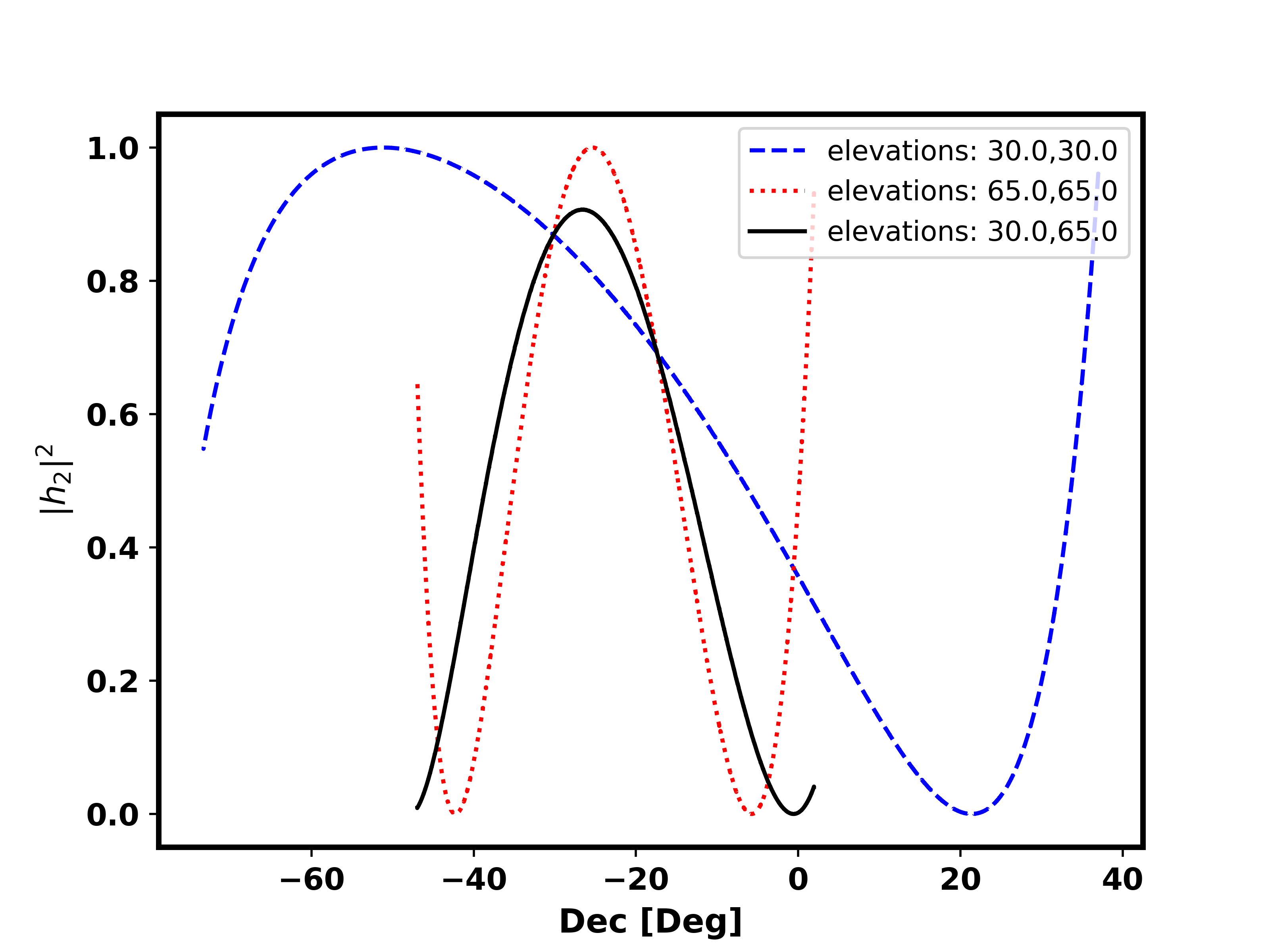}
    \includegraphics[width=3.5in]{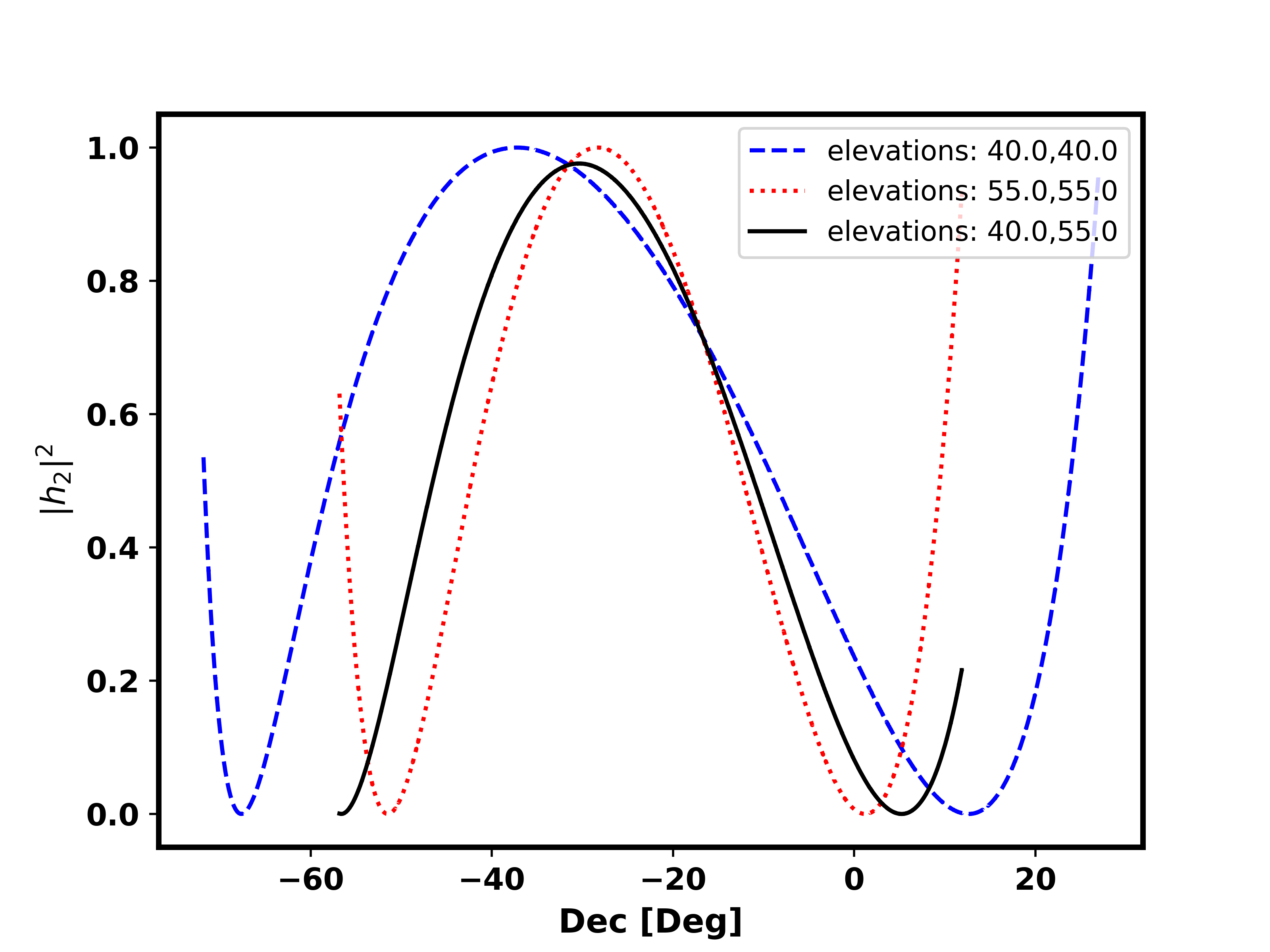}\\
\caption{$|h_{2}|^2$ as a function of declination, comparing 2ERS to single elevation scans for the upper and lower elevations. In each case the black (solid) line is the 2ERS setup, the blue (dashed) line is the single elevation setup for the lower bound and the red (dotted) line is the single elevation setup for the upper bound. The left plot has elevation bounds $(30,65)$ and the right plot has elevation bounds $(40,55)$. The same properties are visible in both: 2ERS has the lowest peak, is better when averaged over the whole visible area, and it is always close to the best per pixel value from a single elevation. However, specific declination ranges can be chosen such that one of the single elevation setups is optimal for the entire declination range. For these specific declination ranges, 2ERS is not a strict lower bound (but it is close), and the respective single elevation scan provides the true bound.}
\label{fig_singleelev}
\end{figure}

We illustrate this in figure \ref{fig_singleelev}. Here we compare the 2ERS setup to setups where only one each of the two limiting elevations are used (but rising and setting are both still used). These plots show that 2ERS has the lowest peak over the whole visible field, is never the worst value over any range, and gives a competitive bound when averaged over the whole visible area. However, specific declination ranges can be chosen such that a single elevation setup is optimal for the entire targeted declination range. For these specific declination ranges, 2ERS is not a strict lower bound on the $h_k$ metrics (but it is close), and the respective single elevation scan provides the true bound.

In more detail, a single elevation scan using the lower elevation from a 2ERS scan typically cuts through the 2ERS peak at a slightly lower value (but is worse for all other declinations). Conversely, a single elevation scan using the upper elevation is typically better for the range of declinations either side of the peak. For example, considering the left hand panel of figure \ref{fig_singleelev}, if one is targeting a declination range of $A^\circ$ to $B^\circ$, then 2ERS would not be a strict lower bound for the average value over the targeted field, and perhaps not a strict lower bound for the peak value, although it does approximate the correct value. If greater precision of the scheduling-independent bound is required, the single elevation scan at the upper limit of $65^\circ$ does provide this bound, and can be used for simple scheduling-independent investigations, which is why we have included single elevations in our results tables. 

\clearpage
\section{C. Peak values}
\label{app_peakvalues}
We show the peak value for different pairs of elevation bounds in table \ref{table_peaks}. 

\begin{table}
 \caption{$|h_2|^2$ worst pixel values for different elevation bounds using 2ERS setup, averaged over the entire sky between declinations $-46.96^\circ$ and $1.94^\circ$).}
\centering
\begin{mytabular}[1.8]{|c|cccccccc|} 
 \hline 
 & $30^\circ$ & $35^\circ$ & $40^\circ$ & $45^\circ$ & $50^\circ$ & $55^\circ$ & $60^\circ$ & $65^\circ$  \\
 \hline 
$30^\circ$ & $1.0$ & $0.99$ & $0.98$ & $0.96$ & $0.94$ & $0.93$ & $0.92$ & $0.91$\\
$35^\circ$ &-& $1.0$ & $0.99$ & $0.98$ & $0.97$ & $0.96$ & $0.94$ & $0.93$\\
$40^\circ$ &-&-& $1.0$ & $1.0$ & $1.0$ & $1.0$ & $1.0$ & $1.0$\\
$45^\circ$ &-&-&-& $1.0$ & $1.0$ & $1.0$ & $1.0$ & $1.0$\\
$50^\circ$ &-&-&-&-& $1.0$ & $1.0$ & $1.0$ & $1.0$\\
$55^\circ$ &-&-&-&-&-& $1.0$ & $1.0$ & $1.0$\\
$60^\circ$ &-&-&-&-&-&-& $1.0$ & $1.0$\\
$65^\circ$ &-&-&-&-&-&-& -& $1.0$ \\
 \hline 
 \end{mytabular} 
 \label{table_peaks}
\end{table}

\bibliographystyle{aasjournal}
\bibliography{ces_note.bib}

\end{document}